\PassOptionsToPackage{table}{xcolor}
\documentclass[a4paper,UKenglish,cleveref, autoref, thm-restate]{lipics-v2021}
\usepackage{booktabs}

\newcommand{\gcL}{\cellcolor{gray!12}}
\newcommand{\gcD}{\cellcolor{gray!35}}
\newcommand{\ob}[1]{\textbf{\color{red}\underline{#1}}}  % overall best
\newcommand{\dup}[1]{\textsuperscript{\textcolor{green!60!black}{+#1}}}  % delta up
\title{Can LLM Coding Assistants Support Emerging Programming Languages? An Empirical Study on Cangjie}

\titlerunning{LLM Coding Assistants on Emerging Languages: A Study on Cangjie}

\author{Junhang Cheng}{Beihang University, Beijing, China}{chengjunhang7@gmail.com}{0009-0001-2254-9198}{}

\author{Fang Liu\footnote{Corresponding author.}}{Beihang University, Beijing, China}{fangliu@buaa.edu.cn}{0000-0002-3905-8133}{}

\author{Jia Li}{Wuhan University, Wuhan, China}{jia.li@whu.edu.cn}{0000-0002-9411-971X}{}

\author{Chengru Wu}{Beihang University, Beijing, China}{23230618@buaa.edu.cn}{0009-0002-0195-0684}{}

\author{Nanxiang Jiang}{Beihang University, Beijing, China}{jiangnx@buaa.edu.cn}{0009-0006-3106-5722}{}

\author{Li Zhang}{Beihang University, Beijing, China}{lily@buaa.edu.cn}{0000-0002-2258-5893}{}

\authorrunning{J. Cheng, F. Liu, J. Li, C. Wu, N. Jiang, and L. Zhang}

\Copyright{Junhang Cheng, Fang Liu, Jia Li, Chengru Wu, Nanxiang Jiang, and Li Zhang}

\ccsdesc[500]{Software and its engineering~Software verification and validation}
\ccsdesc[500]{Computing methodologies~Natural language processing}
\ccsdesc[300]{Software and its engineering~General programming languages}

\keywords{Large language models, Code generation, Code translation, Emerging programming languages, Empirical software engineering, Benchmark, Cangjie}

\category{} %optional, e.g. invited paper

\relatedversiondetails[linktext={arXiv:2603.14501}]{Preprint}{https://arxiv.org/abs/2603.14501}

\supplementdetails[linktext={CangjieBench replication package}, subcategory={Software, Dataset}]{Software}{https://github.com/cjhCoder7/CangjieBench}

\funding{This research is supported by the National Natural Science Foundation of China (Grant No.\ 62302021), the State Key Laboratory of Complex \& Critical Software Environment (Grant No.\ CCSE2025ZX-09), and the Fundamental Research Funds for the Central Universities.}

\nolinenumbers %uncomment to disable line numbering

\EventEditors{Robert Feldt and Maria Paasivaara}
\EventNoEds{2}
\EventLongTitle{20th International Symposium on Empirical Software Engineering and Measurement (ESEM 2026)}
\EventShortTitle{ESEM 2026}
\EventAcronym{ESEM}
\EventYear{2026}
\EventDate{October 8--9, 2026}
\EventLocation{Munich, Germany}
\EventLogo{}
\SeriesVolume{394}
\ArticleNo{44}
\begin{document}

\maketitle

\begin{abstract}
\noindent\textbf{Background.} LLM coding assistants are widely used in development, but evidence about their reliability mostly comes from mainstream languages such as Python and Java. Their behaviour on emerging languages, where public code and documentation are scarce, is far less clear.

\noindent\textbf{Aims.} We ask how capable current LLMs are on Cangjie, an emerging language in the HarmonyOS ecosystem, and which assistive technique is worth its token cost.

\noindent\textbf{Method.} We build CangjieBench, 248 manually translated tasks derived from HumanEval and ClassEval, run under Text-to-Code and Code-to-Code modalities. We evaluate six LLMs under five methods: direct generation, syntax-constrained prompting, retrieval over documentation, retrieval over code, and agent-based workflows. We report Pass@1, compile rate, token cost, and a seven-category failure taxonomy validated by two LLM annotators ($\kappa \geq 0.98$).

\noindent\textbf{Results.} Direct generation is unusable on most models, and over four-fifths of failed samples carry foreign syntax that the Cangjie compiler rejects. A 2{,}146-token syntax reference gives the best accuracy-to-cost trade-off among prompt-based methods. The best agent configuration reaches the highest Pass@1 at 10--120x the token cost of a single prompt, while weaker agents cannot match a syntax-constrained call on the same backbone. A Python reference does not always help: on Code-to-Code it can pull models toward Python idioms and drop compile rates below Text-to-Code.

\noindent\textbf{Conclusions.} No single method dominates both accuracy and token cost. A practical assistant for an emerging language should ship a concise syntax cheat sheet first, escalate to curated retrieval for API-heavy tasks, and reserve agent loops for tasks that need multi-turn repair. The replication package is at \url{https://github.com/cjhCoder7/CangjieBench}.
\end{abstract}

\section{Introduction}
\label{sec:introduction}

Large language models (LLMs) now power IDE completions and agent CLIs~\cite{claude-code,codex-cli,gemini-cli} that generate, translate, and repair code in real projects, but most of the evidence that they do this well rests on a narrow set of languages.
HumanEval~\cite{humaneval} and MBPP~\cite{mbpp} evaluate Python, the benchmarks most often reported for new models~\cite{livecodebench} still centre on Python, Java, and JavaScript, and MultiPL-E~\cite{multipl-e} shows that once evaluation moves to less represented languages the gap between a model's Python score and its score elsewhere can be large.
For languages outside this core, far less is known about whether current models are usable, which assistive technique is worth its token cost, and what the typical failure modes look like.

Emerging languages make the problem worse.
A young language has little public code, its documentation is still being written, and its idioms have not stabilised, so the base model, the retrieval index, and the agent scaffolding all have less material to draw on than on Python or Java, and claims established on Python, such as ``retrieval adds a few points of Pass@1'' or ``agent loops recover most compile errors'', do not automatically carry over and have to be re-measured.

Cangjie is such a language.
Released in 2024 as the native language of the HarmonyOS application ecosystem~\cite{cangjie}, Cangjie is statically typed and memory-safe, with syntax and type rules that differ from both Python and Java.
HarmonyOS now ships on hundreds of millions of devices, so its developers are the intended audience for a Cangjie coding assistant, yet the volume of Cangjie seen at training time is small and the assistant has less prior knowledge of the language than developers expect.
For a team adopting Cangjie today, three questions are open: how capable current models are without help, whether assistive techniques that work for Python transfer, and how the choice between prompt, retrieval, and agent affects accuracy and token cost.

Existing evaluations do not close this gap.
Prior work on low-resource languages has mostly targeted domain-specific languages such as Verilog~\cite{verilogeval} or Solidity~\cite{soleval}, where any performance drop mixes missing domain knowledge with missing language knowledge.
Cross-language benchmarks such as MultiPL-E and HumanEval-XL~\cite{humaneval-xl} translate HumanEval into many targets but do not compare assistive methods or report token cost, and empirical studies of prompt engineering, retrieval-augmented generation~\cite{rag-survey}, and agent-based workflows~\cite{intervenor} are almost entirely set in Python and rarely put these methods on the same accuracy-against-cost axis.
A study of a modern, statically-typed emerging general-purpose language that brings these factors together is missing.

We address that gap with an empirical study organised around four research questions:
\begin{itemize}
  \item \textbf{RQ1 (accuracy).} How do direct generation, syntax-constrained prompting, documentation retrieval, code-example retrieval, and agent-based workflows compare on Cangjie pass rate?
  \item \textbf{RQ2 (token cost).} How much does each method cost in input and output tokens, and how does the cost ranking compare with the pass-rate ranking?
  \item \textbf{RQ3 (transfer).} Is Code-to-Code translation from Python easier than Text-to-Code generation, and does a Python reference induce negative transfer toward Python idioms?
  \item \textbf{RQ4 (failure modes).} What is the distribution of foreign-syntax drift, API misuse, type and mutability errors, duplicate-definition or harness conflicts, runtime exceptions, and logic errors, and how does it shift across methods, models, and tasks?
\end{itemize}

To answer these questions, we build \textbf{CangjieBench}, a 248-task benchmark with 164 function-level problems manually translated from HumanEval and 84 class-level problems from ClassEval~\cite{classeval}, each shipping with a Cangjie reference solution and a harness that compiles the generated code, links it against the reference tests, and runs the result.
On this benchmark we evaluate six recent LLMs (DeepSeek-V3~\cite{deepseek-v3}, ERNIE-4.5~\cite{ernie-4.5}, Kimi-K2~\cite{kimi-k2}, Qwen3-235B-2507 and Qwen3-Coder-480B~\cite{qwen3}, and GPT-5~\cite{gpt5}) under five methods: direct generation, syntax-constrained prompting with a concise rule set, retrieval over Cangjie documentation, retrieval over curated code examples, and an agent-based workflow that reads documentation, runs the compiler, and revises across turns.
Every task runs under two modalities, Text-to-Code and Code-to-Code, and for each run we record Pass@1, compile rate, token usage, and the category of the first observed failure, seeded by manual labelling of 788 failed GPT-5 samples and applied to the full benchmark by two LLM annotators whose agreement we cross-check ($\kappa \geq 0.98$).

Four observations recur across RQ1--RQ4.
Direct generation is unreliable on most models, as code typically fails before the tests run and over four-fifths of failed samples carry foreign syntax the compiler will not accept.
Among prompt-based methods a concise syntax rule set gives the best accuracy-to-cost trade-off and retrieval does not match it once its token cost is counted, while the best agent configuration reaches the highest absolute pass rate but spends an order of magnitude more tokens than any single-shot method.
A Python reference does not always help, and on several cells it pulls the model toward Python idioms so that Code-to-Code compile rate drops below Text-to-Code, a negative-transfer pattern rather than a boost.
Finally, the failure mix shifts with method, as prompt-only runs are dominated by syntax and API errors while agent runs are the only C2C configuration where logic errors reach a non-trivial share, since their failed samples regularly clear the compiler.

Taking the benchmark, the evaluation, and the failure analysis together, this paper makes three contributions.
\begin{itemize}
  \item \textbf{CangjieBench}, a 248-task function- and class-level benchmark for Cangjie, with reference solutions, compile-and-test harnesses, and a public replication package.
  \item A head-to-head evaluation of six LLMs across five methods and two modalities that reports pass rate, compile rate, and token cost separately, making the accuracy-cost trade-off explicit rather than collapsing it into one score.
  \item A failure taxonomy for Cangjie code generation, with per-method distributions and an analysis of where current models break on an emerging language.
\end{itemize}

\section{Background and Related Work}
\label{sec:background}

\noindent\textbf{Low-Resource Language Benchmarks.}
LLMs perform well on high-resource languages~\cite{multipl-e} but drop sharply on low-resource ones, largely due to training-data scarcity.
MultiPL-E extended HumanEval and MBPP to eighteen languages, and more recent benchmarks focus on individual targets such as Verilog~\cite{verilogeval,verilogeval-v2}, Solidity~\cite{soleval}, TeX~\cite{texpert}, VHDL~\cite{vhdl-eval}, shell scripting~\cite{exec-nl2bash}, GPU kernels~\cite{kernelbench}, and SQL~\cite{bird}.
On the translation side, ClassEval-T~\cite{classeval-t} extends class-level evaluation to translation tasks, and MERA Code~\cite{mera} unifies tasks across languages.
A recent survey~\cite{low-resource-survey} organises this space.
Two gaps remain for our study.
First, \textbf{entanglement with domain knowledge}: domain-specific language (DSL) benchmarks mix language novelty with domain constraints, so an observed drop mixes unfamiliar syntax with unfamiliar hardware or contract semantics.
Second, \textbf{data-leakage risk}: languages traditionally labelled low-resource, such as Lua~\cite{evaluating-lua} and R~\cite{r-test}, have long histories and still appear in modern pre-training corpora, weakening contamination controls.
Cangjie addresses both: as a general-purpose language it keeps the domain constant, and its 2024 release keeps pre-training exposure small.

\noindent\textbf{Methods of Code Generation.}
Modern code generation builds on foundation models such as GPT-5, Qwen3, and Gemini~\cite{gemini3}, together with three broad families of assistive methods: \textit{Prompt Engineering}~\cite{code-prompting,doccgen}, \textit{Retrieval-Augmented Generation}~\cite{reposcope}, and \textit{Agentic Frameworks}~\cite{intervenor}.
At the project level, recent work also studies multi-agent collaboration over larger codebases~\cite{project-gen}.
For Python-to-target-language translation specifically, prior work spans parallel corpora such as AVATAR~\cite{avatar}, function-level decomposition~\cite{function-trans}, execution-guided repair~\cite{execoder}, interpretable error correction~\cite{interpretable-trans}, and language-pair benchmarks such as CRUST-Bench for C-to-Rust~\cite{crust-bench}.
Previous studies on low-resource code, however, usually limit evaluation to direct zero-shot or few-shot prompting and rarely report token cost alongside accuracy.
We place all three families on the same footing, with the same tasks, models, and target language, so that differences in accuracy and token cost can be attributed to the method rather than to incomparable setups.

\section{CangjieBench Construction}
\label{sec:benchmark}

\begin{figure}[t]
    \centering
    \includegraphics[alt={Overview of the CangjieBench framework showing three components: dataset construction via manual translation from HumanEval and ClassEval, two evaluation tasks (Text-to-Code and Code-to-Code), and a Docker-based evaluation sandbox that compiles and executes generated Cangjie code.}, width=\linewidth]{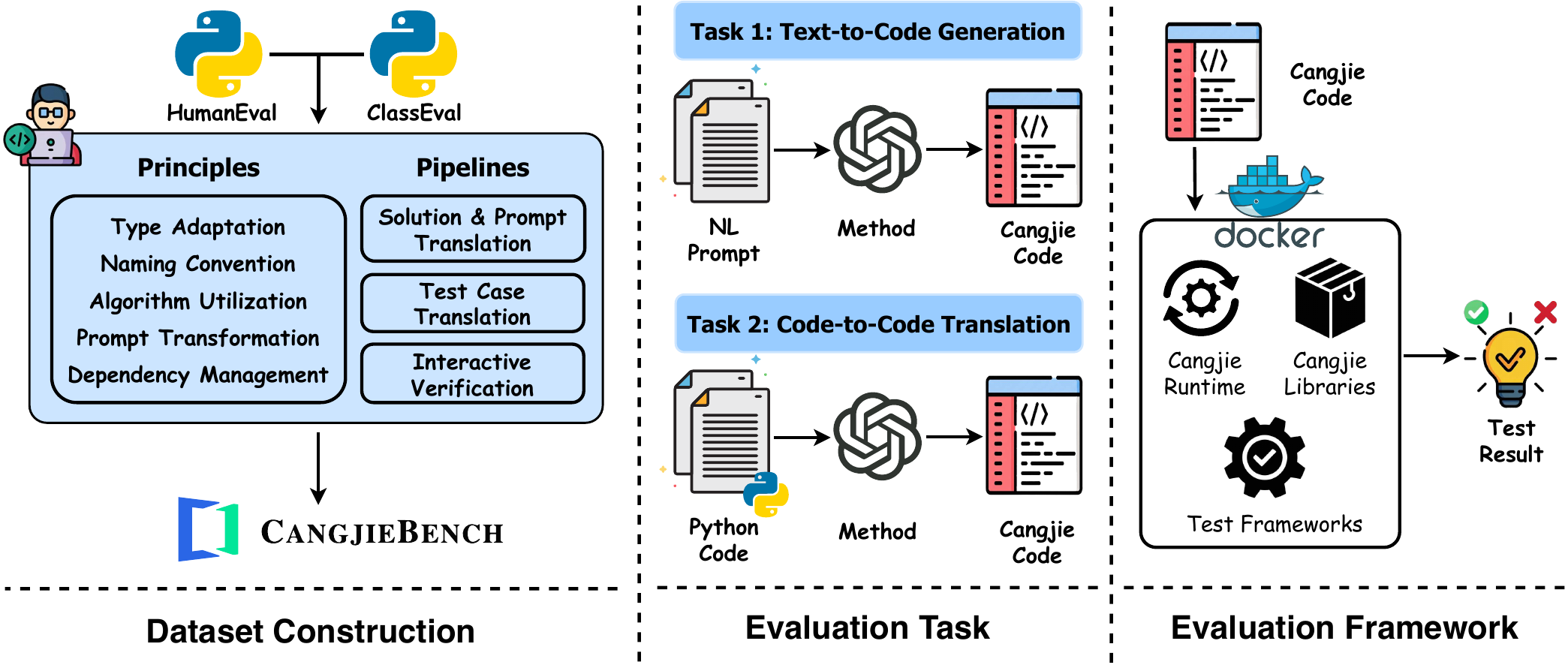}
    \caption{Overview of the CangjieBench framework: dataset construction, evaluation tasks, and evaluation sandbox.}
    \label{fig:benchmark_overview}
\end{figure}

Multilingual suites such as MultiPL-E do not include Cangjie, and DSL benchmarks entangle language novelty with domain knowledge.
We therefore build \textbf{CangjieBench}, which targets Cangjie directly, runs both Text-to-Code generation and Python-to-Cangjie translation on the same problem set, and grades each sample by compiling and executing it on the Cangjie toolchain.
As Figure~\ref{fig:benchmark_overview} shows, it has three components: a manually translated dataset of 248 Cangjie problems, two evaluation tasks under different input modalities, and a framework that compiles and runs each sample inside a Docker sandbox (Section~\ref{sec:data-availability}).

\subsection{Dataset Construction}
\label{sec:benchmark:dataset}

Publicly available Cangjie code is scarce and dominated by official samples a pre-trained model has likely seen, so the usual crawl-and-deduplicate pipeline is not viable.
We instead build CangjieBench by manually translating two widely used Python benchmarks, \textit{HumanEval} and \textit{ClassEval}, covering function-level and class-level problems.
HumanEval~\cite{humaneval} consists of 164 hand-written Python problems, each with a function signature, docstring, and unit tests, and is the de facto standard for function-level generation despite a known bias toward algorithmic tasks and limited API diversity.
ClassEval~\cite{classeval} extends this to the class level with 100 problems that require complete Python classes, testing the multi-method cohesion and cross-method state management absent from function-level benchmarks.
Both datasets are self-contained, one task per file with no cross-file dependencies, unlike repository-level benchmarks such as CoderEval~\cite{codereval} and SWE-bench~\cite{swe-bench}, which keeps the ground truth easy to verify by hand and makes each task small enough that any failure can be attributed to the model's handling of Cangjie rather than to ambiguity about which file to edit.

To keep the translation faithful, we follow a fixed set of construction rules throughout, derived from the authors' experience building Cangjie programs since its 2024 release.
No existing benchmark methodology addresses Cangjie, so rather than adopt one we developed the rules during a pilot translation of roughly 30 tasks and applied them to all 248.
\begin{itemize}
    \item \textbf{Type Adaptation.} We map Python primitives onto their strict Cangjie equivalents (\texttt{int}$\rightarrow$\texttt{Int64}, \texttt{float}$\rightarrow$\texttt{Float64}, \texttt{str}$\rightarrow$\texttt{String}, \texttt{bool}$\rightarrow$\texttt{Bool}) and Python lists and dictionaries onto \texttt{ArrayList} and \texttt{HashMap}. Where a Python signature relies on dynamic typing, we use Cangjie's \texttt{Any} interface and add explicit casts only where the harness needs a concrete value.
    \item \textbf{Naming Convention.} We keep the original \textit{snake\_case} identifiers rather than camel case, which preserves the link to the source task and rules out a trivial failure where the model looks for the wrong identifier.
    \item \textbf{Algorithm Utilization.} Cangjie reference solutions reproduce the control flow of the Python solutions step by step, including branching, loops, and recursion, without substituting a more idiomatic algorithm, since the goal is to grade how well a model expresses a known algorithm in Cangjie rather than the algorithm itself.
    \item \textbf{Prompt Transformation.} HumanEval and ClassEval prompts embed Python fragments such as signatures, docstrings with type hints, and few-shot examples, and we rewrite every fragment into Cangjie form so that the description and any embedded code both refer to the target language, since a mixed-language prompt would leave us unable to tell whether the model was responding to the prompt or patching over the inconsistency.
    \item \textbf{Dependency Management.} We substitute the matching Cangjie standard library for standard Python modules. Tasks depending on third-party Python libraries with no Cangjie equivalent, such as \texttt{sqlite3}, \texttt{PIL}, and \texttt{sklearn}, are excluded rather than reimplemented, since a from-scratch replacement would measure library quality rather than model capability, and Table~\ref{tab:excluded_tasks} lists all 16, but we implement very small utilities such as \texttt{hashlib.md5} so that tasks needing only a hash function stay in.
\end{itemize}

The construction was carried out in three steps over roughly 1.5 months.
First, the first author translated every prompt and reference solution and wrote a matching Cangjie test suite.
Second, two other authors with substantial Cangjie experience independently reviewed each translated problem, solution, and test case, resolving disagreements by discussion and, where needed, by rerunning the tests against alternative references.
Third, we ran every suite against its reference solution and against deliberate perturbations such as wrong return types, off-by-one boundaries, and known-buggy substitutes, confirming that the tests both accept the correct answer and reject common incorrect ones.

After review, CangjieBench contains \textbf{248} tasks: all \textbf{164} HumanEval problems, which map cleanly onto Cangjie's standard library, plus \textbf{84} of the original 100 ClassEval problems, with Table~\ref{tab:excluded_tasks} listing the 16 excluded for Python-specific dependencies.
The two subsets give two granularities: HumanEval probes whether a model can write a single correctly typed routine, while ClassEval asks it to lay out a multi-method class and respect the Cangjie object model across method boundaries.

\begin{table}[t]
    \centering
    \caption{ClassEval tasks excluded from CangjieBench and the Python-specific library responsible for the exclusion.}
    \label{tab:excluded_tasks}
    \small
    \resizebox{\textwidth}{!}{
        \begin{tabular}{lll}
            \toprule
            \textbf{Problem ID} & \textbf{Problem Name} & \textbf{Exclusion Reason} \\
            \midrule
            ClassEval\_14 & \texttt{BookManagementDB} & Relies on \texttt{sqlite3} for database operations \\
            ClassEval\_25 & \texttt{CookiesUtil} & Relies on \texttt{json} for data serialisation \\
            ClassEval\_28 & \texttt{DatabaseProcessor} & Relies on \texttt{sqlite3} for database operations \\
            ClassEval\_34 & \texttt{DocFileHandler} & Relies on \texttt{docx} for Word document processing \\
            ClassEval\_38 & \texttt{ExcelProcessor} & Relies on \texttt{openpyxl} for spreadsheet manipulation \\
            ClassEval\_44 & \texttt{HtmlUtil} & Relies on \texttt{bs4} and \texttt{gensim} for HTML/NLP processing \\
            ClassEval\_45 & \texttt{ImageProcessor} & Relies on \texttt{PIL} for image processing \\
            ClassEval\_50 & \texttt{JSONProcessor} & Relies on \texttt{json} for data serialisation \\
            ClassEval\_52 & \texttt{Lemmatization} & Relies on \texttt{nltk} for NLP processing \\
            ClassEval\_60 & \texttt{MovieTicketDB} & Relies on \texttt{sqlite3} for database operations \\
            ClassEval\_69 & \texttt{PDFHandler} & Relies on \texttt{PyPDF2} for PDF manipulation \\
            ClassEval\_83 & \texttt{StudentDatabaseProcessor} & Relies on \texttt{sqlite3} for database operations \\
            ClassEval\_84 & \texttt{TextFileProcessor} & Relies on \texttt{json} for data serialisation \\
            ClassEval\_92 & \texttt{UserLoginDB} & Relies on \texttt{sqlite3} for database operations \\
            ClassEval\_98 & \texttt{XMLProcessor} & Relies on \texttt{xml} for XML parsing \\
            ClassEval\_99 & \texttt{ZipFileProcessor} & Relies on \texttt{zipfile} for archive management \\
            \bottomrule
        \end{tabular}
    }
\end{table}

\subsection{Evaluation Tasks}
\label{sec:benchmark:tasks}

Every task runs under two modalities that share the same reference solution and test harness.
In \textit{Text-to-Code}, the model receives only a natural-language description and produces a Cangjie implementation from scratch, isolating its ability to generate Cangjie without a per-task anchor.
In \textit{Code-to-Code}, the model additionally receives the corresponding Python reference and must produce a behaviourally equivalent Cangjie implementation, mirroring a team porting an existing Python codebase and letting us measure whether a known-good Python solution helps or instead anchors the model on Python idioms.

We grade correctness by running the per-task Cangjie test suite.
A HumanEval-derived sample passes if the generated function compiles and clears all test cases, and a ClassEval sample must pass every method-level test plus a class-level integration test, so that a class whose methods pass in isolation but fail when composed is still flagged incorrect.

\subsection{Evaluation Framework}
\label{sec:benchmark:framework}

The evaluation framework is a Docker-based sandbox that pins the Cangjie toolchain, standard library, and test runner to a single image and executes each sample in a fresh container.
Pinning lets us compare timing and error messages across models, methods, and machines on equal terms, and isolating each run protects the host from agent-generated code the model has only partially understood.
For every sample the harness records whether compilation succeeded, whether all tests passed, the compiler and runtime stderr, and the wall-clock time, the raw signals on which the failure analysis in Section~\ref{sec:results} is built.

\section{Methods}
\label{sec:methods}

\begin{figure}[t]
    \centering
    \includegraphics[alt={Diagram of the four method families evaluated on CangjieBench: Direct Generation (baseline), Syntax-Constrained Prompting (adds a concise grammar reference), RAG-Docs and RAG-Code (retrieval-augmented generation over documentation and code examples respectively), and Agent-based workflows (multi-turn compiler-feedback loop).}, width=\linewidth]{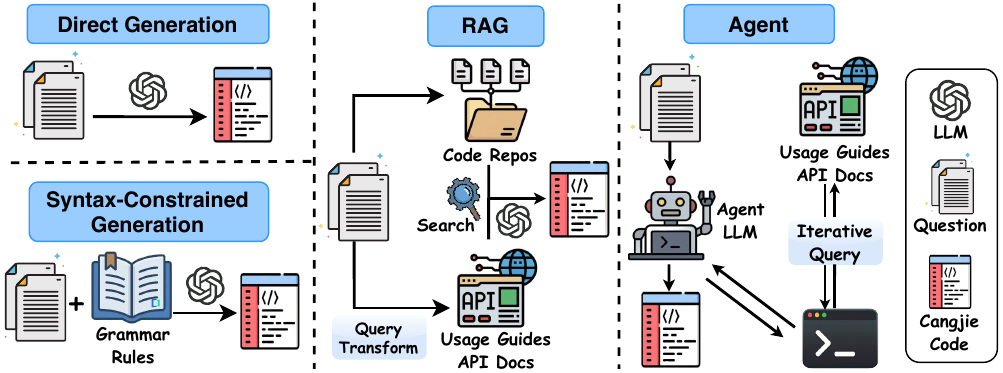}
    \caption{The four method families evaluated on CangjieBench. Direct generation is the baseline. The other three families add Cangjie-specific context in different ways.}
    \label{fig:methods}
\end{figure}

We exclude training methods, since the goal is not to engineer a Cangjie-specialised model but to ask how far an off-the-shelf foundation model can be pushed on an emerging language using techniques a team would normally have at hand.
We evaluate four method families common in practice, summarised in Figure~\ref{fig:methods}: \textit{Direct Generation}, \textit{Syntax-Constrained Prompting}, \textit{Retrieval-Augmented Generation} with two variants (RAG-Docs and RAG-Code), and \textit{Agent-based workflows}.

\subsection{Method Families}
\label{sec:methods:families}

\noindent\textbf{Direct Generation.}
The model receives the natural-language problem description, plus the Python reference solution under Code-to-Code, wrapped in a fixed prompt template with no grammar reference, no retrieved snippets, and no tools.

\noindent\textbf{Syntax-Constrained Prompting.}
Since most of the LLMs we evaluate have seen little Cangjie at training time, a common engineering response is to ship a small ``language pack'', a syntax cheat-sheet injected into every prompt, and we prepend such a reference to the direct-generation template.

\noindent\textbf{Retrieval-Augmented Generation.}
A more involved response gives the assistant access to a Cangjie corpus at inference time and pastes a few retrieved snippets into context.
We evaluate two variants matching the two corpora a typical assistant would index, official documentation (RAG-Docs) and crawled code (RAG-Code), and for an emerging language both corpus and retrieval signal are weaker than for Python, so whether this pays off is an open question.

\noindent\textbf{Agent-based Workflow.}
A more open-ended response, similar to how a new developer learns an unfamiliar language, lets the assistant read documentation, write code, run the compiler, and revise its output across several turns, and we want to know what such a loop costs in tokens and buys in pass rate on a language with thin pre-training coverage.

\subsection{Configuration}
\label{sec:methods:configuration}

\noindent\textbf{Syntax reference.}
The grammar reference used by Syntax-Constrained Prompting covers 20 categories: program structure, variables, numeric types, strings and characters, the boolean type, arrays and dynamic arrays, hash maps, hash sets, sorting, tuples, conditional statements, loop statements, the option type, match expressions, lambda expressions, input and output, functions, classes, interfaces, and reserved keywords.
It totals 2{,}146 GPT-5 tokens, well below every model's context limit.

\noindent\textbf{RAG corpora and retriever.}
Both RAG variants use Okapi BM25 with top-$k=3$.
The \textit{Docs} corpus is the Cangjie Developer Manual and standard-library API reference from the official repository, with HTML stripped and text split on section headers, and because problem statements and API reference rarely share surface vocabulary, we add a query-transformation step in which the model first extracts technical keywords such as library and API method names, which we issue to BM25 in place of the raw description.
The \textit{Code} corpus is Cangjie source crawled from public GitCode repositories before 10 November 2025, filtered to files at least 80\,\% Cangjie, parsed with the Cangjie AST tool, and split into 30{,}341 functions and 31{,}483 class definitions, and here we issue the prompt directly to BM25 because it already contains code-shaped material such as signatures and type hints.

\noindent\textbf{Agent harnesses.}
We use three CLI agent harnesses without modification: \textbf{Codex CLI} version 0.63.0, \textbf{Qwen Code CLI} version 0.2.3, and \textbf{iFlow CLI} version 0.4.8.
All three can read files, execute shell commands, and inspect shell output, and their built-in retrieval differs, itself a method-level variable.
Each agent receives the problem description, the direct-generation template, and read access to the same Cangjie documentation tree used by RAG-Docs, kept under Git and rolled back after every interaction step, and terminates when its harness reports completion or hits the step budget.

\section{Results}
\label{sec:results}

\begin{table}[t]
\centering
\caption{Pass@1 (\%) and Compile rate (\%) on CangjieBench. \colorbox{gray!12}{Light gray} columns indicate Text-to-Code (T2C), and \colorbox{gray!35}{dark gray} columns indicate Code-to-Code (C2C). The best result per method is in \textbf{bold}, and the overall best across all methods is \ob{red and underlined}. Avg.\ averages over HumanEval (164 tasks) and ClassEval (84 tasks) weighted by task count.}
\label{tab:main_results}
\resizebox{\textwidth}{!}{%
\begin{tabular}{ll cccc cccc cccc}
\toprule
\multirow{2.5}{*}{\textbf{Method}} & \multirow{2.5}{*}{\textbf{Subject}} & \multicolumn{4}{c}{\textbf{HumanEval}} & \multicolumn{4}{c}{\textbf{ClassEval}} & \multicolumn{4}{c}{\textbf{Avg.}} \\
\cmidrule(lr){3-6} \cmidrule(lr){7-10} \cmidrule(lr){11-14}
 & & \multicolumn{2}{c}{Pass@1} & \multicolumn{2}{c}{Compile} & \multicolumn{2}{c}{Pass@1} & \multicolumn{2}{c}{Compile} & \multicolumn{2}{c}{Pass@1} & \multicolumn{2}{c}{Compile} \\
 & & T2C & C2C & T2C & C2C & T2C & C2C & T2C & C2C & T2C & C2C & T2C & C2C \\
\midrule
\multirow{6}{*}{\textbf{Direct}}
 & DeepSeek-V3 & \gcL 2.4 & \gcD 2.4 & \gcL 3.0 & \gcD 3.0 & \gcL 1.2 & \gcD 1.2 & \gcL 1.2 & \gcD 1.2 & \gcL 2.0 & \gcD 2.0 & \gcL 2.4 & \gcD 2.4 \\
 & ERNIE-4.5 & \gcL 3.0 & \gcD 4.3 & \gcL 4.3 & \gcD 4.9 & \gcL 0.0 & \gcD 1.2 & \gcL 0.0 & \gcD 1.2 & \gcL 2.0 & \gcD 3.2 & \gcL 2.8 & \gcD 3.6 \\
 & Kimi-K2 & \gcL \textbf{20.7} & \gcD \textbf{21.3} & \gcL \textbf{23.8} & \gcD \textbf{23.8} & \gcL \textbf{4.8} & \gcD \textbf{8.3} & \gcL \textbf{7.1} & \gcD \textbf{8.3} & \gcL \textbf{15.3} & \gcD \textbf{16.9} & \gcL \textbf{18.1} & \gcD \textbf{18.5} \\
 & Qwen3 & \gcL 3.7 & \gcD 3.7 & \gcL 4.3 & \gcD 4.3 & \gcL 1.2 & \gcD 2.4 & \gcL 1.2 & \gcD 2.4 & \gcL 2.8 & \gcD 3.2 & \gcL 3.2 & \gcD 3.6 \\
 & Qwen3-Coder & \gcL 3.7 & \gcD 7.3 & \gcL 4.3 & \gcD 7.9 & \gcL 1.2 & \gcD 1.2 & \gcL 1.2 & \gcD 1.2 & \gcL 2.8 & \gcD 5.2 & \gcL 3.2 & \gcD 5.6 \\
 & GPT-5 & \gcL 6.7 & \gcD 7.9 & \gcL 7.3 & \gcD 8.5 & \gcL 1.2 & \gcD 3.6 & \gcL 1.2 & \gcD 3.6 & \gcL 4.8 & \gcD 6.5 & \gcL 5.2 & \gcD 6.9 \\
\midrule
\multirow{6}{*}{\textbf{\shortstack[l]{Syntax-\\Constrained}}}
 & DeepSeek-V3 & \gcL 40.9 & \gcD 42.7 & \gcL 47.6 & \gcD 44.5 & \gcL 9.5 & \gcD 6.0 & \gcL 16.7 & \gcD 6.0 & \gcL 30.2 & \gcD 30.2 & \gcL 37.1 & \gcD 31.5 \\
 & ERNIE-4.5 & \gcL 36.6 & \gcD 32.3 & \gcL 39.0 & \gcD 35.4 & \gcL 1.2 & \gcD 4.8 & \gcL 2.4 & \gcD 6.0 & \gcL 24.6 & \gcD 23.0 & \gcL 26.6 & \gcD 25.4 \\
 & Kimi-K2 & \gcL 55.5 & \gcD \textbf{53.0} & \gcL 62.2 & \gcD \textbf{56.1} & \gcL 11.9 & \gcD 14.3 & \gcL 22.6 & \gcD 15.5 & \gcL 40.7 & \gcD \textbf{39.9} & \gcL 48.8 & \gcD \textbf{42.3} \\
 & Qwen3 & \gcL 53.0 & \gcD 45.1 & \gcL 57.3 & \gcD 47.6 & \gcL 13.1 & \gcD 10.7 & \gcL 22.6 & \gcD 14.3 & \gcL 39.5 & \gcD 33.5 & \gcL 45.6 & \gcD 36.3 \\
 & Qwen3-Coder & \gcL 42.7 & \gcD 48.8 & \gcL 47.6 & \gcD 51.8 & \gcL 16.7 & \gcD 11.9 & \gcL 22.6 & \gcD 13.1 & \gcL 33.9 & \gcD 36.3 & \gcL 39.1 & \gcD 38.7 \\
 & GPT-5 & \gcL \textbf{64.0} & \gcD 44.5 & \gcL \textbf{67.1} & \gcD 45.1 & \gcL \textbf{23.8} & \gcD \textbf{25.0} & \gcL \textbf{40.5} & \gcD \textbf{31.0} & \gcL \textbf{50.4} & \gcD 37.9 & \gcL \textbf{58.1} & \gcD 40.3 \\
\midrule
\multirow{6}{*}{\textbf{\shortstack[l]{RAG\\(Code)}}}
 & DeepSeek-V3 & \gcL 15.2 & \gcD 14.0 & \gcL 16.5 & \gcD 15.9 & \gcL 2.4 & \gcD 4.8 & \gcL 3.6 & \gcD 4.8 & \gcL 10.9 & \gcD 10.9 & \gcL 12.1 & \gcD 12.1 \\
 & ERNIE-4.5 & \gcL 13.4 & \gcD 17.7 & \gcL 15.9 & \gcD 18.9 & \gcL 4.8 & \gcD 3.6 & \gcL 6.0 & \gcD 3.6 & \gcL 10.5 & \gcD 12.9 & \gcL 12.5 & \gcD 13.7 \\
 & Kimi-K2 & \gcL 30.5 & \gcD 26.8 & \gcL 33.5 & \gcD 31.7 & \gcL \textbf{8.3} & \gcD 9.5 & \gcL \textbf{13.1} & \gcD 9.5 & \gcL 23.0 & \gcD 21.0 & \gcL 26.6 & \gcD 24.2 \\
 & Qwen3 & \gcL 12.8 & \gcD 7.3 & \gcL 13.4 & \gcD 7.9 & \gcL 3.6 & \gcD 7.1 & \gcL 3.6 & \gcD 7.1 & \gcL 9.7 & \gcD 7.3 & \gcL 10.1 & \gcD 7.7 \\
 & Qwen3-Coder & \gcL 23.2 & \gcD 17.1 & \gcL 25.0 & \gcD 18.3 & \gcL 3.6 & \gcD 6.0 & \gcL 7.1 & \gcD 7.1 & \gcL 16.5 & \gcD 13.3 & \gcL 19.0 & \gcD 14.5 \\
 & GPT-5 & \gcL \textbf{47.0} & \gcD \textbf{45.1} & \gcL \textbf{49.4} & \gcD \textbf{47.0} & \gcL \textbf{8.3} & \gcD \textbf{10.7} & \gcL \textbf{13.1} & \gcD \textbf{13.1} & \gcL \textbf{33.9} & \gcD \textbf{33.5} & \gcL \textbf{37.1} & \gcD \textbf{35.5} \\
\midrule
\multirow{6}{*}{\textbf{\shortstack[l]{RAG\\(Docs)}}}
 & DeepSeek-V3 & \gcL 27.4 & \gcD 21.3 & \gcL 33.5 & \gcD 22.6 & \gcL 6.0 & \gcD 8.3 & \gcL 9.5 & \gcD 9.5 & \gcL 20.2 & \gcD 16.9 & \gcL 25.4 & \gcD 18.1 \\
 & ERNIE-4.5 & \gcL 11.6 & \gcD 11.6 & \gcL 13.4 & \gcD 12.8 & \gcL 2.4 & \gcD 3.6 & \gcL 3.6 & \gcD 3.6 & \gcL 8.5 & \gcD 8.9 & \gcL 10.1 & \gcD 9.7 \\
 & Kimi-K2 & \gcL 32.3 & \gcD 30.5 & \gcL 34.8 & \gcD \textbf{34.1} & \gcL \textbf{7.1} & \gcD \textbf{16.7} & \gcL 11.9 & \gcD \textbf{16.7} & \gcL 23.8 & \gcD \textbf{25.8} & \gcL 27.0 & \gcD \textbf{28.2} \\
 & Qwen3 & \gcL 11.6 & \gcD 10.4 & \gcL 12.8 & \gcD 12.2 & \gcL 1.2 & \gcD 4.8 & \gcL 1.2 & \gcD 4.8 & \gcL 8.1 & \gcD 8.5 & \gcL 8.9 & \gcD 9.7 \\
 & Qwen3-Coder & \gcL 19.5 & \gcD 17.7 & \gcL 22.6 & \gcD 18.9 & \gcL 3.6 & \gcD 8.3 & \gcL 6.0 & \gcD 9.5 & \gcL 14.1 & \gcD 14.5 & \gcL 16.9 & \gcD 15.7 \\
 & GPT-5 & \gcL \textbf{36.0} & \gcD \textbf{31.1} & \gcL \textbf{37.2} & \gcD 32.3 & \gcL \textbf{7.1} & \gcD 10.7 & \gcL \textbf{15.5} & \gcD \textbf{16.7} & \gcL \textbf{26.2} & \gcD 24.2 & \gcL \textbf{29.8} & \gcD 27.0 \\
\midrule
\multirow{4}{*}{\textbf{Agent}}
 & Kimi-K2 (iFlow CLI) & \gcL 39.0 & \gcD 48.2 & \gcL 44.5 & \gcD 51.8 & \gcL 17.9 & \gcD 22.6 & \gcL 26.2 & \gcD 27.4 & \gcL 31.9 & \gcD 39.5 & \gcL 38.3 & \gcD 43.5 \\
 & Qwen3-Coder (iFlow CLI) & \gcL 29.9 & \gcD 29.9 & \gcL 32.3 & \gcD 33.5 & \gcL 4.8 & \gcD 14.3 & \gcL 9.5 & \gcD 17.9 & \gcL 21.4 & \gcD 24.6 & \gcL 24.6 & \gcD 28.2 \\
 & Qwen3-Coder (Qwen Code CLI) & \gcL 25.0 & \gcD 23.8 & \gcL 30.5 & \gcD 27.4 & \gcL 7.1 & \gcD 7.1 & \gcL 8.3 & \gcD 8.3 & \gcL 19.0 & \gcD 18.1 & \gcL 23.0 & \gcD 21.0 \\
 & GPT-5 (Codex CLI) & \gcL \ob{84.8} & \gcD \ob{86.6} & \gcL \ob{87.2} & \gcD \ob{87.8} & \gcL \ob{32.1} & \gcD \ob{58.3} & \gcL \ob{67.9} & \gcD \ob{65.5} & \gcL \ob{66.9} & \gcD \ob{77.0} & \gcL \ob{80.6} & \gcD \ob{80.2} \\
\bottomrule
\end{tabular}}
\end{table}

\subsection{Experimental Setup}
\label{sec:results:setup}

\noindent\textbf{Subjects.}
The open-weights subjects are DeepSeek-V3 (671B parameters), ERNIE-4.5 (300B), Kimi-K2-0905 (1T), Qwen3-235B-A22B-Instruct-2507 (235B), and Qwen3-Coder-480B, and the closed-source subject is GPT-5.
We restrict the selection to models above 200B parameters because smaller models bottomed out at near-zero Pass@1 in pilot runs, leaving no headroom to separate the methods, and the six we keep span two families of general-purpose models and two code-specialised ones, so any method effect we report is not tied to a single training recipe.
The four agent configurations are Codex CLI / GPT-5, Qwen Code CLI / Qwen3-Coder, iFlow CLI / Qwen3-Coder, and iFlow CLI / Kimi-K2-0905, using the three CLI harnesses of Section~\ref{sec:methods:configuration}, chosen because they are the harnesses a Cangjie developer can install and run today.

\noindent\textbf{Decoding protocol.}
All cells use greedy decoding with \texttt{temperature}~$=0$ and \texttt{max\_tokens}~$=8192$, drawing a single sample per task, with open-source models served through the SiliconFlow API and GPT-5 through the OpenAI API.
This removes sampling variance from the comparison but forgoes the smoothing of Pass@$k$ with $k>1$, so we add a Pass@10 robustness check under syntax-constrained prompting in Section~\ref{sec:discussion}.
For agent runs we cap each task at the harness default step budget and count a run as failed if it exhausts the budget without a compiling submission.

\noindent\textbf{Metrics.}
\textit{Pass@1} is the fraction of tasks whose single sample compiles and passes every test, and \textit{Compile rate} is the fraction that compile regardless of tests, so their gap separates logic-only failures from cases where no compilable Cangjie program was produced.
\textit{Token usage} sums input and output tokens over the 248 tasks, aggregating every turn for agent runs, and \textit{Failure category distribution} reports the share of unsuccessful samples in each of the seven failure classes of Section~\ref{sec:results:rq4}.
We report Pass@1 and Compile rate as percentages weighted by task count across the two subsets, and the per-cell records behind every number are in the replication package (Section~\ref{sec:data-availability}).

\subsection{RQ1: How do the four method families compare on pass rate?}
\label{sec:results:rq1}

In Table~\ref{tab:main_results} the four method families do not order themselves on a single accuracy axis, as the three findings below make precise.

\textbf{Direct generation fails at the compile step, not the logic step.}
Without Cangjie-specific scaffolding, five of the six base models score under 7\,\% Pass@1 in either modality, only Kimi-K2 breaks 10\,\% at 15.3\,\% T2C and 16.9\,\% C2C on average, and GPT-5, the strongest model elsewhere, reaches only 4.8\,\% on Direct T2C.
Compile rate sits within about 3\,pp of Pass@1 on every Direct cell, so a failing model usually fails before the tests run, unable to produce a program the toolchain accepts.

\textbf{Syntax-Constrained Prompting dominates the prompt-based methods.}
On T2C average, every base model gains 22--46\,pp over Direct once given the grammar reference: GPT-5 climbs from 4.8\,\% to 50.4\,\%, Kimi-K2 from 15.3\,\% to 40.7\,\%, and Qwen3 from 2.8\,\% to 39.5\,\%.
Both RAG variants help less, as averaged over T2C and C2C GPT-5 reaches 33.7\,\% under RAG-Code, 25.2\,\% under RAG-Docs, and 44.2\,\% under Syntax-Constrained, with the same ordering on the open-weights subjects.
That a 2{,}146-token grammar reference outperforms BM25 retrieval over megabytes of Cangjie source shows that \textbf{fixing the syntax model matters more than supplying example code}, since retrieved examples cannot fix syntax the model has not yet seen.

\textbf{Agent reaches the highest pass rate, but only with a strong backbone.}
Codex CLI on GPT-5 averages 66.9\,\% T2C and 77.0\,\% C2C, the best on every subset and modality and the only Agent configuration that dominates Syntax-Constrained on its own backbone, while the other three sit in the 18--40\,\% range, comparable to or below Syntax-Constrained on the same backbones.
The agentic loop amplifies what the backbone can already do, but does not lift a weak backbone past a stronger one running with the syntax reference.

\begin{figure}[t]
    \centering
    \begin{minipage}{0.49\linewidth}
        \centering
        \includegraphics[alt={Stacked bar chart showing per-task token usage of Qwen3-Coder on Text-to-Code across five methods on a log scale, split into input (prompt) and output (completion) percentages with absolute totals labelled above each bar.}, width=\linewidth]{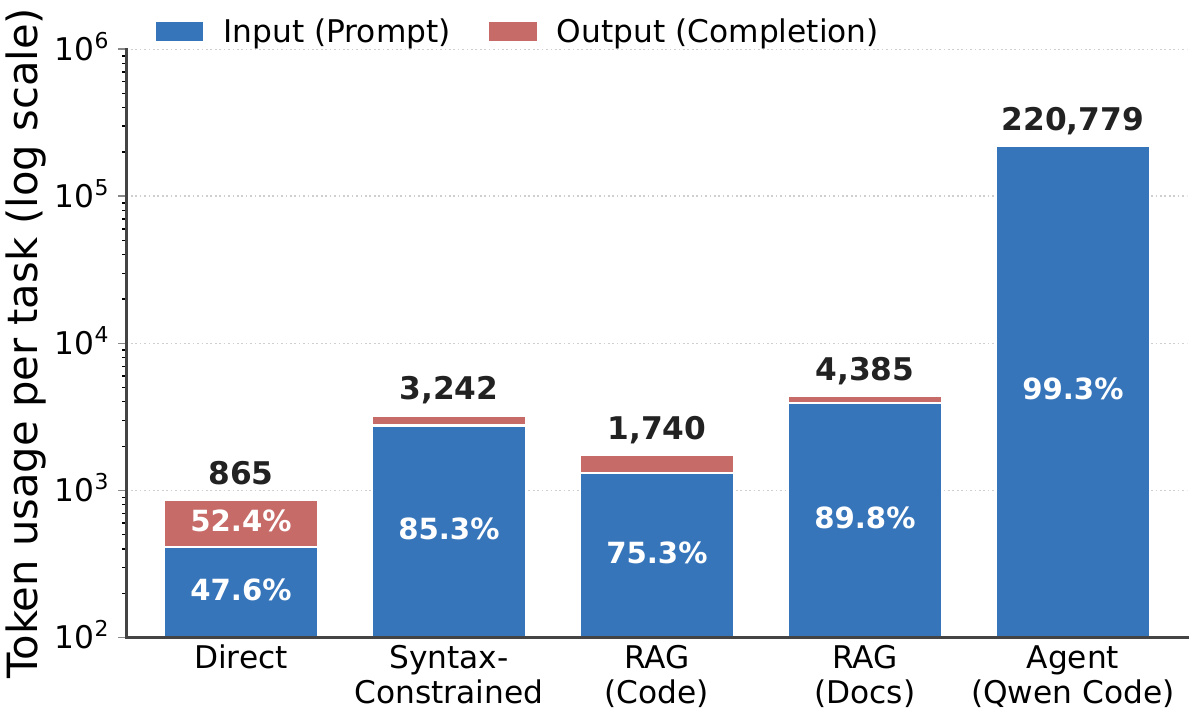}\\
        {\small (a) Text-to-Code}
    \end{minipage}\hfill
    \begin{minipage}{0.49\linewidth}
        \centering
        \includegraphics[alt={Stacked bar chart showing per-task token usage of Qwen3-Coder on Code-to-Code across five methods on a log scale, split into input (prompt) and output (completion) percentages with absolute totals labelled above each bar.}, width=\linewidth]{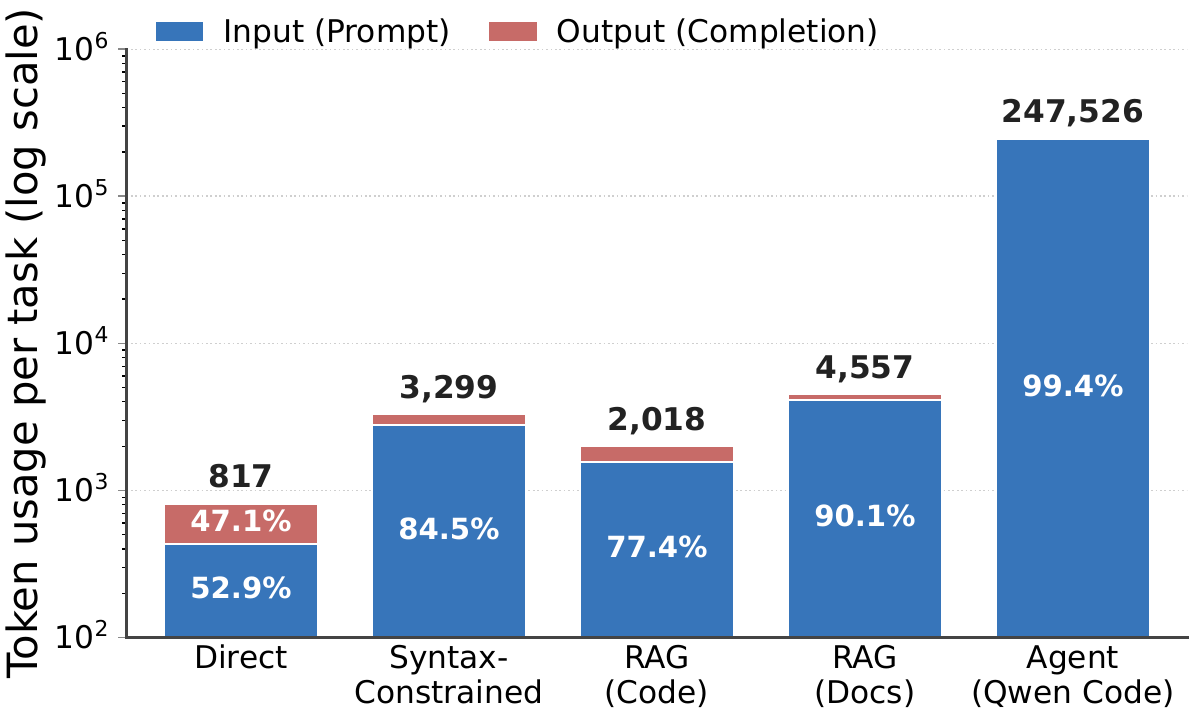}\\
        {\small (b) Code-to-Code}
    \end{minipage}
    \caption{Per-task token usage of Qwen3-Coder across the five methods on (a) Text-to-Code and (b) Code-to-Code. Total tokens are on a log scale, and each bar is split into Input (Prompt) and Output (Completion) percentages, with the absolute total printed above.}
    \label{fig:qwen_token}
\end{figure}

\subsection{RQ2: How much does each method cost in tokens?}
\label{sec:results:rq2}

Figure~\ref{fig:qwen_token} reports Qwen3-Coder's per-task token usage across all five methods, spanning two orders of magnitude.
Direct generation consumes roughly 1k tokens per task, Syntax-Constrained about 3.3k once the 2{,}146-token grammar header is added, and the two RAG variants between about 1.7k and 4.6k, while Agent under iFlow CLI jumps to 34--36k and Agent under Qwen Code CLI to 221--248k, roughly 220--250x Direct.
Output share falls from about half of Direct's tokens to below 3\,\% for both agent runs, so the assistant spends almost all of its budget reading documentation and prior turns rather than writing code.

\begin{figure}[t]
    \centering
    \includegraphics[alt={Bar chart of Pass@1 per 1000 input tokens for Kimi-K2, Qwen3-Coder, and GPT-5 on Text-to-Code across five methods. Higher bars indicate better accuracy-to-cost efficiency. Direct generation on Kimi-K2 is the most efficient; agent methods are the least efficient.}, width=\linewidth]{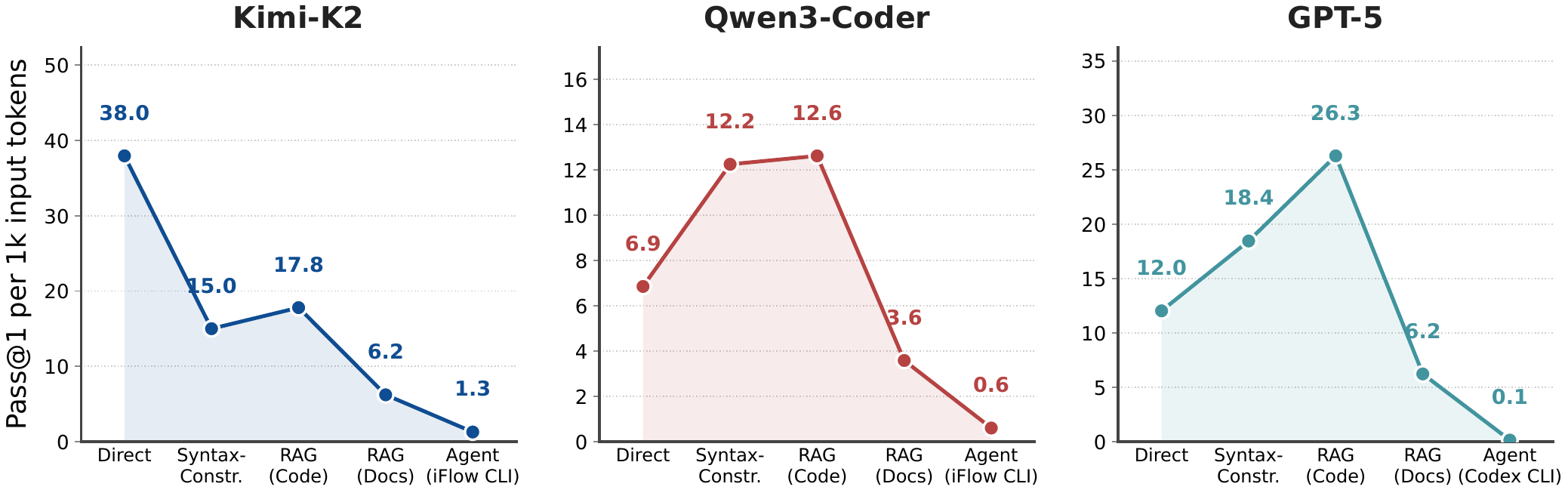}\\
    {\small (a) Text-to-Code}\\[0.5em]
    \includegraphics[alt={Bar chart of Pass@1 per 1000 input tokens for Kimi-K2, Qwen3-Coder, and GPT-5 on Code-to-Code across five methods. Pattern mirrors the Text-to-Code chart with agent methods showing the lowest efficiency.}, width=\linewidth]{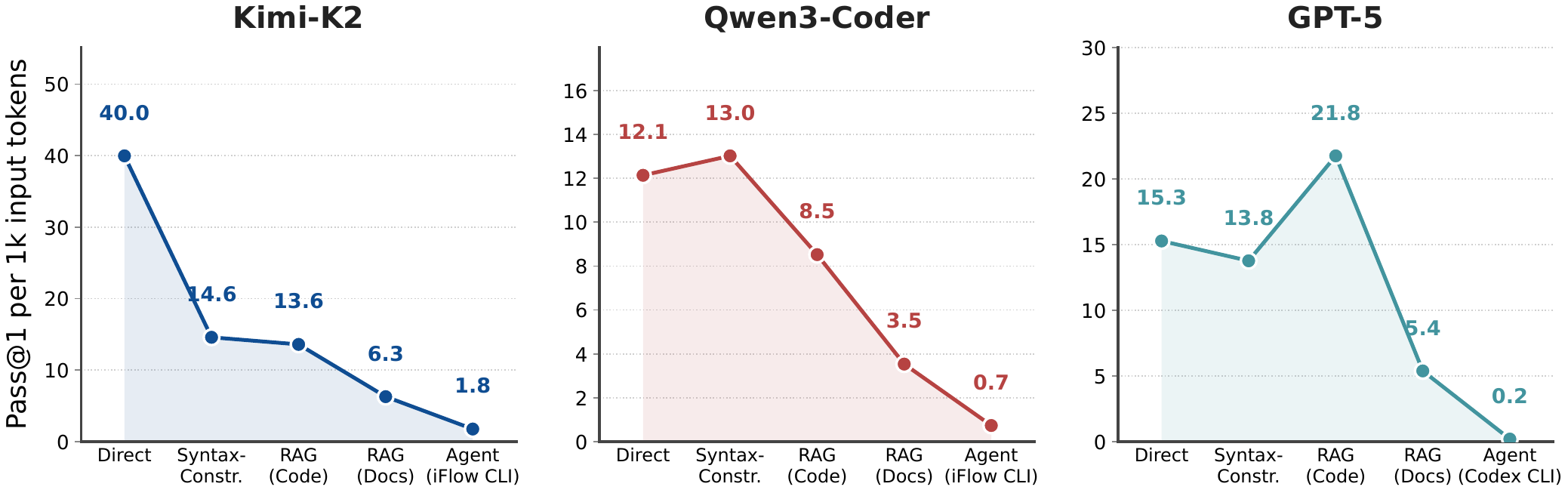}\\
    {\small (b) Code-to-Code}
    \caption{Pass@1 per 1{,}000 input tokens for Kimi-K2, Qwen3-Coder, and GPT-5 on (a) Text-to-Code and (b) Code-to-Code. Pass@1 and input tokens are task-count-weighted over HumanEval (164) and ClassEval (84), matching Table~\ref{tab:main_results}. Higher is better.}
    \label{fig:pass_per_1k_token}
\end{figure}

\textbf{Cost reverses the accuracy ranking.}
Figure~\ref{fig:pass_per_1k_token} plots Pass@1 per 1{,}000 input tokens, on which a method that doubles accuracy at tenfold the tokens loses.
On Kimi-K2, Direct is the most efficient cell at around 40 Pass@1 per 1k tokens on both modalities despite scoring only 15--17\,\% Pass@1, since a single 500-token prompt yields a little correct code very cheaply.
On Qwen3-Coder and GPT-5, where Direct yields almost no passing samples, Syntax-Constrained and RAG-Code lead, and GPT-5 RAG-Code on T2C reaches 26.3 Pass@1 per 1k tokens, more than 4x RAG-Docs at 6.2 on the same backbone.

\textbf{The Agent ceiling collapses on the per-token axis.}
Codex CLI on GPT-5 sets the absolute Pass@1 ceiling, but Figure~\ref{fig:pass_per_1k_token} shows it consumes those gains at 0.14 Pass@1 per 1k tokens on T2C and 0.20 on C2C, tens to nearly two hundred times below the prompt-based methods on the same backbone, and the two iFlow configurations sit at 0.6--1.8.
Method choice therefore becomes a question of token budget rather than absolute accuracy: when tokens are unconstrained Codex CLI on GPT-5 is the right choice, but under any practical budget \textbf{Syntax-Constrained Prompting is the dominant strategy on every backbone we evaluate}, reaching on Qwen3-Coder 33.9\,\% T2C and 36.3\,\% C2C at 10--75x lower cost than the same backbone in either Agent harness.

\subsection{RQ3: Does a Python reference help or hurt?}
\label{sec:results:rq3}

\begin{figure}[t]
    \centering
    \includegraphics[alt={Heatmap of Delta Pass@1 (Code-to-Code minus Text-to-Code, in percentage points) for each method and model combination. Red cells indicate negative transfer where the Python reference hurts performance; blue cells indicate positive transfer. Syntax-Constrained and RAG-Code cluster on the red side; Direct, RAG-Docs, and Agent configurations cluster on the blue side.}, width=0.95\linewidth]{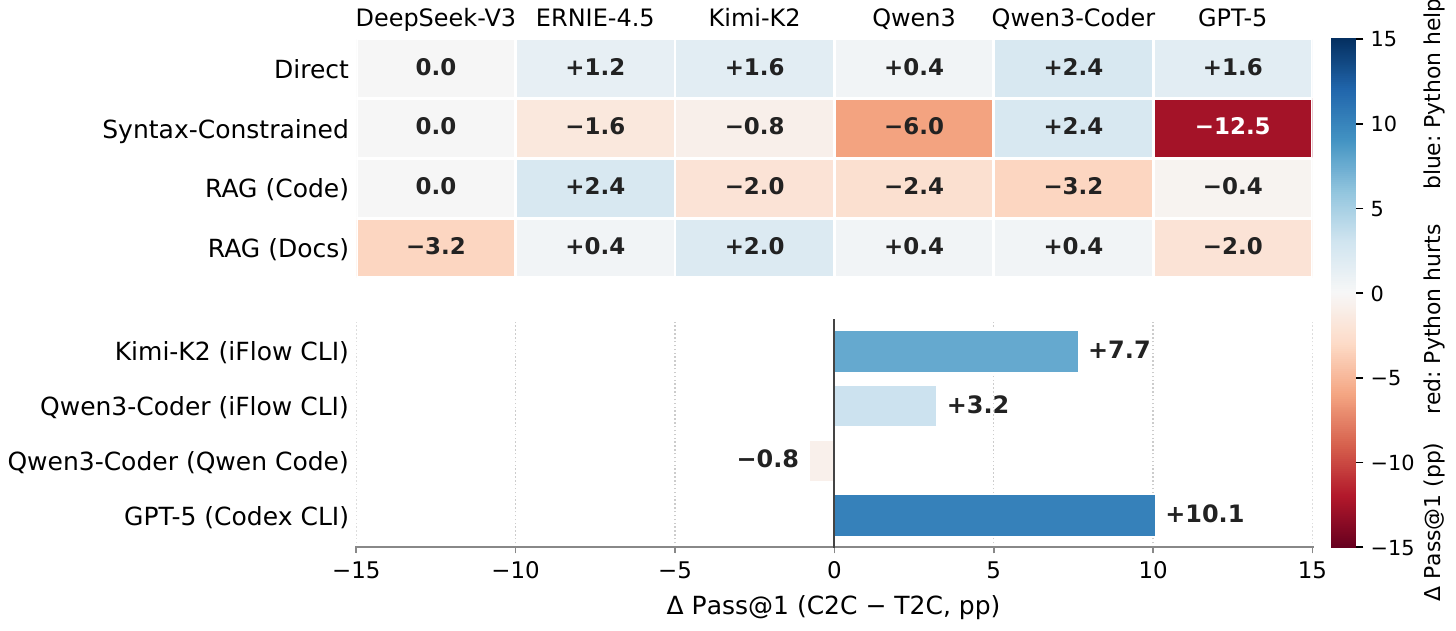}
    \caption{$\Delta$ Pass@1 (C2C $-$ T2C, percentage points), task-count-weighted over HumanEval and ClassEval. Negative (red) means a Python reference hurts on Code-to-Code, and positive (blue) means it helps. Top: prompt-based methods $\times$ base models. Bottom: each of the four agent configurations.}
    \label{fig:python_reference_delta}
\end{figure}

\textbf{Code-to-Code is not uniformly easier than Text-to-Code.}
Averaged across all six base models and all prompt-based methods, C2C and T2C are within about 1\,pp on HumanEval and T2C is slightly ahead on ClassEval.
The Python reference neither reliably helps nor hurts, since its direction depends on the method and its effect on the average is small next to the method-level variance in Figure~\ref{fig:python_reference_delta}, which shows $\Delta$ Pass@1 in percentage points, C2C minus T2C, for every cell, with the Cangjie-anchored prompt-based methods on the red side where the reference hurts and Direct, RAG-Docs, and the Agent configurations on the blue side where it helps.

The most striking case is GPT-5 under Syntax-Constrained on HumanEval, where Pass@1 drops from 64.0\,\% T2C to 44.5\,\% C2C, a 19.5\,pp regression on the same tasks with the Python solution available, and the average column drops 12.5\,pp.
Compile rate moves with it, from 67.1\,\% to 45.1\,\%, so the regression is a syntax issue rather than a logic one, with the reference pulling GPT-5 toward Python idioms that do not survive the Cangjie compiler.

The pattern splits cleanly by method type: under \textit{Syntax-Constrained} and \textit{RAG-Code}, which put Cangjie syntax or code in the prompt, T2C beats C2C on 4 of 6 base models on the average column, so negative transfer concentrates where the model already has Cangjie-shaped material to anchor on.
Two apparent exceptions confirm the mechanism rather than break it.
Syntax-Constrained on Qwen3-Coder shows a slight positive $\Delta$ of about +2.4\,pp, the same direction it takes under Direct, because this code-specialised model carries structure over from the Python reference faster than it takes on the Python idioms the compiler rejects.
RAG-Docs on GPT-5 shows a negative $\Delta$ of about $-$2.0\,pp despite no Cangjie code in the prompt, because retrieved documentation fragments introduce inconsistent Cangjie API calls that the reference compounds.
Under \textit{Direct} and \textit{RAG-Docs}, by contrast, the reference helps on most subjects, since with no Cangjie scaffold to anchor on the Python solution at least gives a structural skeleton.

The Agent rows show a sharply different pattern: three of the four configurations score higher under C2C than T2C on the average column, by 10.1\,pp for Codex on GPT-5, 7.7\,pp for iFlow on Kimi-K2, and 3.2\,pp for iFlow on Qwen3-Coder, with only Qwen Code on Qwen3-Coder flat at $-$0.8\,pp, and the gap is largest on ClassEval, where Codex on GPT-5 jumps from 32.1\,\% to 58.3\,\%.
The multi-turn loop absorbs the negative-transfer pressure, since the extra turns let the agent revise Python-specific syntax that would land verbatim in a single-shot prompt, so \textbf{negative transfer is a property of the consuming method, not of the Python reference itself.}

\subsection{RQ4: Where do the failures come from?}
\label{sec:results:rq4}

We labelled every failed sample with one of seven error categories.

\noindent\textbf{Taxonomy.}
The taxonomy was derived through open coding of 788 failed GPT-5 samples on T2C across all five methods.
We selected GPT-5 because it reaches the widest range of failure modes, whereas weaker models stall almost entirely on foreign-syntax errors, and we seeded on T2C because the model cannot attribute failures to the Python reference, so each error traces to a gap in its Cangjie knowledge.
The first author grouped samples by recurring surface patterns and merged or split categories until no large cluster remained unlabelled, producing six categories, with a seventh, \textit{Timeout}, added for runs that do not finish within the evaluation budget.

\noindent\textbf{Annotation procedure.}
For each failed sample we supplied the LLM annotator with the compiler or runtime error output, the first failing error line, and the category descriptions with one short example each, and instructed it to assign exactly one label, report a confidence level, give a one-sentence rationale, and quote the primary evidence.

\noindent\textbf{Annotator selection.}
We chose glm-5.1 and kimi-k2.6 as annotators because both have strong Chinese-language understanding, which matters as Cangjie documentation and error messages often mix English identifiers with Chinese descriptions, and they come from different organisations, Zhipu AI and Moonshot AI, so agreement across two independently trained models is stronger evidence of label reliability than two runs of one model.
Treating glm-5.1 as primary and kimi-k2.6 as a cross-check, applying both to the 788 manually annotated samples reproduced the manual labels on almost every sample, so we extended LLM labelling to the full benchmark.

The seven categories are \textit{Syntax (foreign)} for invalid syntax mirroring another language, \textit{API/library} for nonexistent identifiers or wrong call shapes, \textit{Type/mutability} for type or mutability mismatches, \textit{Duplicate def.} for redeclarations or harness conflicts, \textit{Runtime exception} for uncaught exceptions, \textit{Logic error} for wrong test outputs, and \textit{Timeout} for runs that did not finish. Table~\ref{tab:failure_distribution} reports the failure mix per method and modality.

\begin{table}[t]
\centering
\caption{Failure distribution per method and modality, primary annotator glm-5.1. Cells show the share of failed samples in each category. Avg.\ aggregates failed samples across methods within each modality. The last row gives the total failure count for each column.}
\label{tab:failure_distribution}
\resizebox{\textwidth}{!}{%
\begin{tabular}{l r@{\hskip 4pt}r r@{\hskip 4pt}r r@{\hskip 4pt}r r@{\hskip 4pt}r r@{\hskip 4pt}r r@{\hskip 4pt}r}
\toprule
\multirow{2}{*}{\textbf{Failure category}} & \multicolumn{2}{c}{\textbf{\shortstack{~\\Direct}}} & \multicolumn{2}{c}{\makebox[0pt][c]{\textbf{\shortstack{Syntax-\\Constrained}}}} & \multicolumn{2}{c}{\textbf{\shortstack{RAG\\(Code)}}} & \multicolumn{2}{c}{\textbf{\shortstack{RAG\\(Docs)}}} & \multicolumn{2}{c}{\textbf{\shortstack{~\\Agent}}} & \multicolumn{2}{c}{\textbf{Avg.}} \\
\cmidrule(lr){2-3} \cmidrule(lr){4-5} \cmidrule(lr){6-7} \cmidrule(lr){8-9} \cmidrule(lr){10-11} \cmidrule(lr){12-13}
 & T2C & C2C & T2C & C2C & T2C & C2C & T2C & C2C & T2C & C2C & T2C & C2C \\
\midrule
Syntax (foreign) & \textbf{83.9} & \textbf{82.0} & 30.0 & 27.2 & \textbf{56.6} & \textbf{59.1} & \textbf{60.9} & \textbf{61.0} & \textbf{46.1} & \textbf{50.4} & \textbf{58.8} & \textbf{58.7} \\
API/library & 12.4 & 14.0 & \textbf{34.6} & \textbf{42.4} & 32.5 & 30.4 & 26.4 & 26.6 & 30.0 & 30.0 & 26.0 & 27.5 \\
Type/mutability & 2.6 & 3.2 & 16.3 & 16.8 & 8.2 & 8.7 & 4.6 & 6.8 & 13.3 & 13.7 & 7.9 & 8.9 \\
Duplicate def. & 0.0 & 0.1 & 8.9 & 10.2 & 0.0 & 0.0 & 4.5 & 3.7 & 0.0 & 0.0 & 2.6 & 2.7 \\
Runtime exc. & 0.4 & 0.5 & 0.7 & 0.8 & 0.6 & 0.7 & 0.4 & 0.5 & 1.2 & 1.0 & 0.6 & 0.7 \\
Logic error & 0.6 & 0.1 & 9.2 & 2.6 & 2.0 & 1.0 & 3.1 & 1.4 & 9.4 & 4.7 & 4.0 & 1.6 \\
Timeout & 0.0 & 0.0 & 0.2 & 0.0 & 0.1 & 0.0 & 0.1 & 0.0 & 0.0 & 0.2 & 0.1 & 0.0 \\
\midrule
\textit{Total failures} & 1{,}414 & 1{,}396 & 944 & 990 & 1{,}229 & 1{,}243 & 1{,}238 & 1{,}243 & 647 & 597 & 5{,}472 & 5{,}469 \\
\bottomrule
\end{tabular}}
\end{table}

\textbf{Foreign-syntax errors dominate, and only the grammar reference cuts them.}
Direct generation fails almost entirely on syntax, with about 83\,\% of its failed samples on both T2C and C2C being \textit{Syntax (foreign)}, code that borrows Python, Rust, Java, or Swift constructs the Cangjie compiler rejects, such as \texttt{if x > 0:} written without parentheses, \texttt{for i in range(n)} loops, and \texttt{|acc, x|} closure syntax, matching the roughly 3\,pp Pass@1-to-Compile gap on every Direct cell.
The pattern is consistent across the weaker models, which almost never emit a construct Cangjie accepts once it diverges from Python, so their low Pass@1 reflects a parser they cannot satisfy rather than an algorithm they cannot find.
The other methods address this with mixed success: Syntax-Constrained cuts foreign-syntax failures to about 28--30\,\%, but RAG-Code and RAG-Docs still leave 56--61\,\% despite putting Cangjie tokens in the prompt, because retrieval supplies surface text without overriding the model's syntactic priors, and Agent sits at 46--51\,\% where the multi-turn loop revises some surface errors.

\textbf{Where the rest of the failures shift.}
Once Syntax-Constrained removes most foreign syntax, the mass shifts onto API/library at 34--42\,\% and Type/mutability at around 16\,\%, because the grammar reference teaches the parser-level shape of Cangjie but not the standard library or type rules, so models write valid Cangjie that calls methods Cangjie lacks or mixes incompatible numeric types.
The C2C column spikes at 42.4\,\% on API/library versus 34.6\,\% on T2C, consistent with RQ3, as the Python reference seeds Python-style API names Cangjie rejects.
About 9--10\,\% of Syntax-Constrained failures are Duplicate def., where the model declares its own \texttt{main} and conflicts with the harness, and Logic error reaches a non-trivial level only under Agent on C2C, at 4.7\,\%, the only method where many failed samples reach the test phase with code the toolchain accepts.

\begin{figure}[t]
    \centering
    \includegraphics[alt={Two 7-by-7 inter-annotator confusion matrices comparing glm-5.1 and kimi-k2.6 labels for (a) Text-to-Code and (b) Code-to-Code. Both matrices are nearly diagonal, indicating near-perfect agreement with Cohen's kappa of 0.989 and 0.981 respectively.}, width=\linewidth]{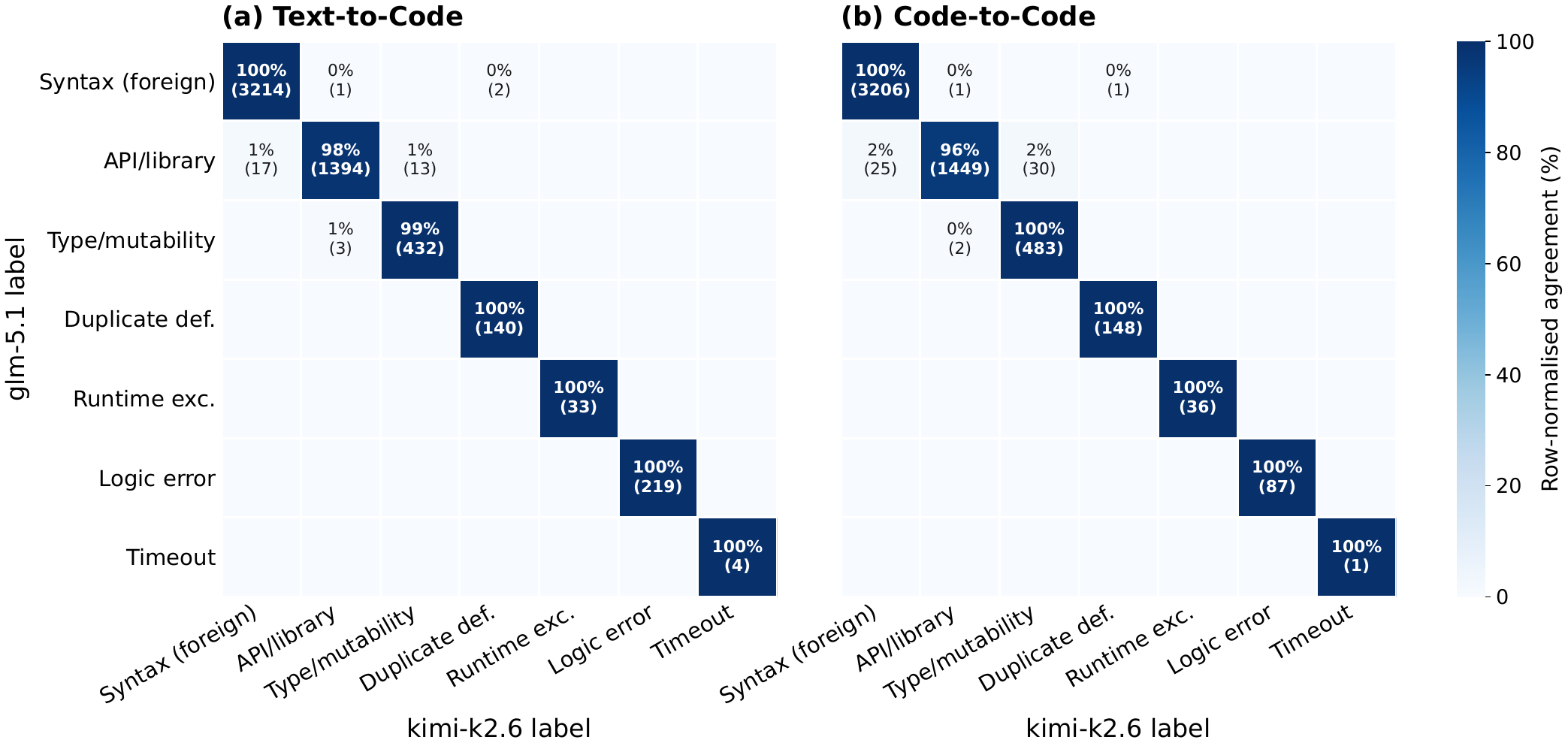}
    \caption{Inter-annotator confusion matrix between glm-5.1 (rows) and kimi-k2.6 (columns) for (a) T2C and (b) C2C. Each cell shows the row-normalised percentage and the absolute count.}
    \label{fig:error_label_confusion}
\end{figure}

\textbf{Cross-annotator validation.}
Figure~\ref{fig:error_label_confusion} reports the 7$\times$7 confusion matrix between glm-5.1 and kimi-k2.6. Both matrices are almost entirely diagonal, with raw agreement 99.34\,\% on T2C ($\kappa = 0.989$) and 98.92\,\% on C2C ($\kappa = 0.981$) and no off-diagonal pair above 2\,\% of any row, so the percentages in Table~\ref{tab:failure_distribution} do not depend on the choice of annotator.

\section{Discussion}
\label{sec:discussion}

For Cangjie, the bottleneck on every task is parser-level rather than reasoning-level, since direct generation fails almost entirely on syntax and the Compile-to-Pass@1 gap is at most about 3\,pp on every Direct cell.
The model already knows the algorithm but cannot yet write a program the toolchain accepts, so what separates the four method families is how each closes parser distance to Cangjie and at what token cost.

\subsection{Spending the Agent's Budget on Repeated Syntax Sampling}
\label{sec:discussion:pass10}

Section~\ref{sec:results:rq2} shows an agent run costs 10--120x a single Syntax-Constrained call for a single-digit Pass@1 gain, so an obvious alternative is to keep the cheap method and draw several samples instead of one long agent trace.
We test this with ten Syntax-Constrained samples per Text-to-Code task, scored under task-level Pass@10, and match it in Table~\ref{tab:syntax_pass10_agent} against the Pass@1 of the same backbone wrapped in each agent harness.

\begin{table}[t]
\centering
\caption{Repeated syntax-guided sampling against single-run agents on Text-to-Code. \textit{Syntax} reports task-level Pass@10 from ten Syntax-Constrained samples per task, and \textit{Agent} reports Pass@1 from one CLI run. {\dup{N.N}} shows the pp gain of Syntax Pass@10 over Agent Pass@1. \textit{Tokens} is the input-plus-output sum per task. Same backbone within each row.}
\label{tab:syntax_pass10_agent}
\small
\resizebox{\textwidth}{!}{%
\begin{tabular}{ll rrrr rrrr}
\toprule
\multirow{3}{*}{\textbf{Backbone}} &
\multirow{3}{*}{\textbf{Agent harness}} &
\multicolumn{4}{c}{\textbf{HumanEval}} &
\multicolumn{4}{c}{\textbf{ClassEval}} \\
\cmidrule(lr){3-6} \cmidrule(lr){7-10}
 & & \multicolumn{2}{c}{Syntax} & \multicolumn{2}{c}{Agent} & \multicolumn{2}{c}{Syntax} & \multicolumn{2}{c}{Agent} \\
 & & Pass@10 & Tokens & Pass@1 & Tokens & Pass@10 & Tokens & Pass@1 & Tokens \\
\midrule
Kimi-K2 & iFlow CLI & 84.1\,\dup{45.1} & 27.1k & 39.0 & 24.2k & 28.6\,\dup{10.7} & 36.6k & 17.9 & 27.5k \\
Qwen3-Coder & iFlow CLI & 64.0\,\dup{34.1} & 28.1k & 29.9 & 36.1k & 19.0\,\dup{14.2} & 42.2k & 4.8 & 36.2k \\
Qwen3-Coder & Qwen Code CLI & 64.0\,\dup{39.0} & 28.1k & 25.0 & 224.8k & 19.0\,\dup{11.9} & 42.2k & 7.1 & 212.9k \\
GPT-5 & Codex CLI & 96.3\,\dup{11.5} & 46.4k & 84.8 & 391.7k & 42.9\,\dup{10.8} & 82.8k & 32.1 & 618.2k \\
\bottomrule
\end{tabular}}
\end{table}

\textbf{Repeated sampling beats the matched agent in every cell, on both subsets.}
On Kimi-K2 the gap is 45.1\,pp on HumanEval at roughly iFlow's token cost, and on Qwen3-Coder against Qwen Code CLI it is 39.0\,pp on HumanEval at one-eighth of the agent's budget.
Even GPT-5 under Codex, the strongest cell in Table~\ref{tab:main_results}, gains 11.5\,pp on HumanEval and 10.8\,pp on ClassEval by switching to parallel syntax-guided samples within roughly one-eighth of Codex's budget.

This is not an unconditional argument against agents.
Pass@10 assumes a test harness that can pick the passing sample, and without tests ten samples are ten guesses.
With tests, the budget pays off more on alternative implementations than on long tool-use traces, because most failures the agent repairs are surface syntax errors a fresh draft would not have made, and repeated sampling is parallelisable, so its wall-clock cost can be below an agent run that waits on each compiler call.
The edit-and-rerun loop thus earns its tokens only when the failure mix rewards serial repair rather than a fresh draft.

\subsection{Why Retrieval Helps Less Than Expected}
\label{sec:discussion:rag}

Both RAG variants trail Syntax-Constrained Prompting on all base models, and failure analysis sharpens the point, since 56--61\% of RAG failures are foreign-syntax failures (Table~\ref{tab:failure_distribution}) even with Cangjie tokens in the prompt.
Retrieved context is plentiful but rarely corrective.

\begin{figure}[t]
    \centering
    \includegraphics[alt={Two side-by-side RAG failure case studies. Left: on the all-prefixes task, RAG-Docs retrieves ArrayList documentation but the task needs string slicing; the model writes a substring call that Cangjie does not expose, causing a compiler error. Right: on the sum-product task, RAG-Code retrieves syntactically valid but irrelevant Cangjie snippets; the model still emits a Python-style while loop without parentheses.}, width=\linewidth]{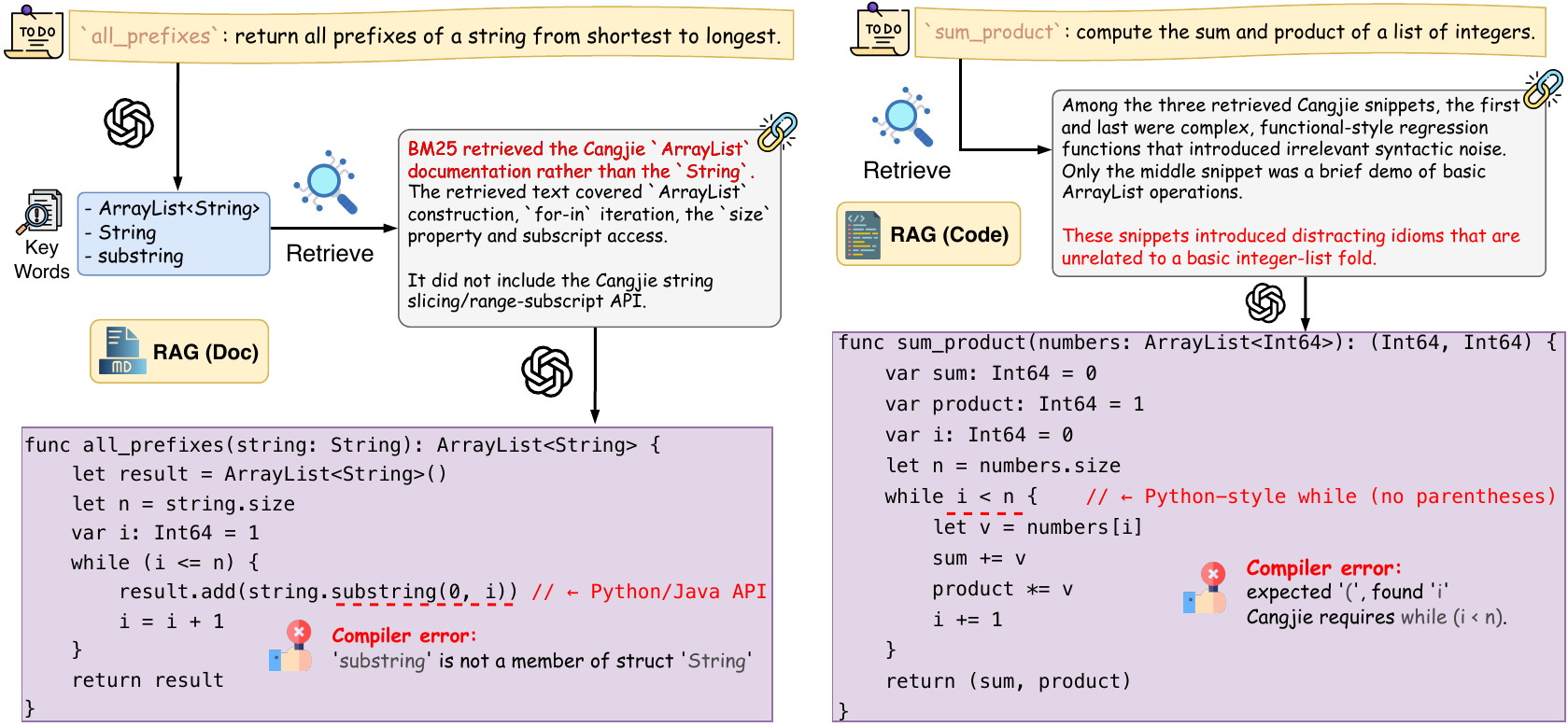}
    \caption{Two representative RAG failures. On \texttt{all\_prefixes}, RAG-Docs returns \texttt{ArrayList} documentation when the task needs Cangjie string slicing, and the model writes \texttt{substring} (an API Cangjie does not expose), which the compiler rejects. On \texttt{sum\_product}, RAG-Code returns syntactically valid but task-irrelevant snippets, and the model still emits \texttt{while i < n} without parentheses.}
    \label{fig:rag_case_study}
\end{figure}

Figure~\ref{fig:rag_case_study} illustrates two cases where retrieval returns documentation or source unrelated to the task and the model falls back to a foreign API name or a Python-style loop the compiler rejects, so retrieval produces context but not the context that would prevent the failure.

For an emerging language this matters more than for Python, because with a thinner corpus a noisy index can no longer compensate by sheer volume, and two failure modes drive the foreign-syntax share even when the prompt is full of Cangjie text.
The first is vocabulary mismatch, where a natural-language description and a Cangjie source file share almost no surface tokens, so a BM25 query for ``slice the prefix of a string'' rarely surfaces the slicing example a developer would open.
The second is the small-corpus problem, where among only a few tens of thousands of crawled functions many APIs appear once or not at all, so the snippet the model needs is often absent.
A more useful index would be small and compiler-checked, organised by API signature and paired with short examples for the constructs that most often break, loops, string handling, option types, and mutability, so that retrieval aims at the most likely compiler error rather than enlarging the prompt with plausible but inert text.
Until such an index exists, the RAG cells in Table~\ref{tab:main_results} are best read as adding tokens that do not move parser distance, not as a real alternative to the syntax reference.

\subsection{A Staged Workflow for Emerging-Language Assistants}
\label{sec:discussion:method_choice}

No single ranking covers both accuracy and token cost, so the practical question is which method to reach for at each stage of a workflow.
Direct Generation is best read as a measurement of how much Cangjie a model already knows rather than a deployment choice.
Syntax-Constrained Prompting is the appropriate first stage, since it dominates both RAG variants on every base model, addresses the largest failure class, and adds only a 2{,}146-token header, and when a test harness is available repeated syntax-guided sampling extends it to recover most of the agent's headline accuracy at a fraction of the budget (Section~\ref{sec:discussion:pass10}).
Curated retrieval, indexed by signatures rather than free text, is worth its tokens on API-heavy tasks where the syntax reference leaves API/library failures unaddressed, and only after a Syntax-Constrained attempt has surfaced a concrete API gap.
Agents are best reserved for tasks that genuinely require interaction, such as inspecting files, preserving multi-method class structure, or repairing Python-specific drift across turns, where Section~\ref{sec:results:rq3} shows them earning their cost on Code-to-Code.
The staging is therefore an escalation triggered by the failure class the previous step cannot fix.

\section{Threats to Validity}
\label{sec:threats}

\noindent\textbf{Construct validity.}
Pass@1 and Compile rate measure whether the toolchain accepts the program and whether it passes the tests we wrote, not whether the code is idiomatic or maintainable in a real codebase.
We mitigate this by writing the tests against the same reference solutions we translated and reviewing each suite against deliberate perturbations so that obvious wrong answers are rejected.
Token cost counts only API tokens rather than wall-clock time, and agent runs that interleave shell calls may incur additional costs, so agent figures are a lower bound on the true gap.

\noindent\textbf{Internal validity.}
Every cell uses greedy decoding with a single sample, so the comparison is sensitive to sampling noise, and we report a ten-sample Pass@10 robustness check under Syntax-Constrained Prompting in Section~\ref{sec:discussion:pass10}.
The failure categories are produced by an LLM annotator rather than a human, so we cross-checked them against a second LLM and against 788 manually inspected GPT-5 failures before extending the labels to the rest of the benchmark.
The translation rules, taxonomy seed set, and category definitions are themselves methodological choices, so we derived the rules from pilot work rather than post-hoc, seeded the taxonomy on the strongest model across all five methods for coverage, and checked that the categories are jointly exhaustive and mutually exclusive on the inspection set.
Our Cangjie reference solutions and tests were written by hand and could contain bugs, so two additional authors reviewed every problem, and we verified that each suite accepts the reference solution and rejects deliberately broken variants.

\noindent\textbf{External validity.}
We evaluate one emerging language, six base models, and three agent harnesses, so any quantitative claim is tied to that subject set.
The qualitative pattern, that direct generation fails at the parser before the algorithm and that a small grammar reference closes most of the gap, holds because Cangjie is statically typed with limited public code; we expect it to transfer to other emerging general-purpose languages in the same early-adoption phase, but not to dynamically typed languages or DSLs, where the failure surface is shaped by domain semantics rather than parser distance.
Tasks come from HumanEval and ClassEval, which are algorithm-level rather than repository-level, so our results say little about long-context Cangjie work such as multi-file edits or build configuration, where an agent's ability to navigate a project may matter far more than single-file syntax.
Pinning the toolchain and CLI versions means absolute pass rates are a snapshot; the ordering across methods is wide enough that it is unlikely to reverse under minor version changes.

\section{Conclusion}
\label{sec:conclusion}

We asked whether current LLM coding assistants can support an emerging programming language, and the answer from CangjieBench is that they can, but only when the method is matched to where the model actually fails.
For the six models we evaluated, the failure on Cangjie is not one of reasoning: direct generation fails almost entirely on syntax, a short syntax reference of about two thousand tokens closes most of that gap, and retrieval helps less than its prompt size suggests because a BM25 index over public Cangjie code rarely returns the snippet that would prevent the next compiler error.
The best agent configuration reaches the highest absolute Pass@1, but only at ten to one hundred and twenty times the cost of a single prompt, while weaker agents do not match a Syntax-Constrained call on the same backbone.
For a team adopting an emerging language today, the implication is to ship a syntax cheat sheet first, add curated retrieval once it is tailored to the language, and reserve an agent loop for tasks that need file inspection or multi-turn repair.
The same staging also guards against the negative transfer we observed under Code-to-Code, where a Python reference pulls a single-shot prompt toward idioms the Cangjie compiler rejects and only the multi-turn loop reliably repairs them.
We expect the qualitative pattern, that syntax is the barrier before semantics, to hold for other statically typed languages in their early-adoption phase, with the thresholds shifting as public code and documentation accumulate.
We release CangjieBench, the harness, and the failure taxonomy, validated by two LLM annotators with $\kappa \geq 0.98$, so that the same comparison can be repeated as models and tooling evolve.

\section{Data Availability}
\label{sec:data-availability}

All data used in this study are available under the MIT licence at \url{https://github.com/cjhCoder7/CangjieBench}.
The repository contains five components, cross-referenced below to the sections that use them.

\begin{enumerate}
  \item \textbf{CangjieBench dataset} (Section~\ref{sec:benchmark:dataset}).
  The 248-task benchmark ships as two JSONL files, one holding the 164 function-level HumanEval tasks and one the 84 class-level ClassEval tasks, with each entry giving the task identifier, Cangjie prompt, reference solution, and test harness, alongside the per-task Cangjie and Python source files.

  \item \textbf{Evaluation sandbox} (Section~\ref{sec:benchmark:framework}).
  A Docker image packaging the Cangjie~1.0 compiler, runtime, standard libraries, and standard extensions (stdx), exposing three REST endpoints (\texttt{/evaluate\_humaneval}, \texttt{/evaluate\_classeval}, and \texttt{/run\_code}) for automated compilation and testing. We used the same image to produce every cell in Table~\ref{tab:main_results}.

  \item \textbf{Per-cell evaluation records} (Sections~\ref{sec:results:rq1}--\ref{sec:results:rq3}).
  Per model, method, and task we provide the raw generation, the compile/pass status, compiler and runtime stderr, and the input and output token counts, with per-cell summaries and the Pass@10 data from Section~\ref{sec:discussion:pass10}.

  \item \textbf{Failure analysis} (Section~\ref{sec:results:rq4}).
  Both annotators, glm-5.1 and kimi-k2.6, labelled every failed sample with one of the seven categories, each record giving the category, annotator confidence, the compiler or runtime evidence, and a short rationale, with per-cell distribution summaries.

  \item \textbf{Prompt templates and method resources} (Section~\ref{sec:methods}).
  We provide the four prompt templates for Direct, Syntax-Constrained, RAG-Docs, and RAG-Code together with the agent prompt for both modalities, the 2{,}146-token Cangjie syntax reference used by Syntax-Constrained Prompting, and the two RAG corpora, namely the official Cangjie documents for RAG-Docs and the crawled GitCode source for RAG-Code.
\end{enumerate}

\noindent\textbf{Generative AI tool usage.}
We used large language models, glm-5.1 and kimi-k2.6, as automated annotators to label failure categories across the full benchmark, as described in Section~\ref{sec:results:rq4}.
All model-generated labels were cross-validated against each other and against a manually annotated seed set.

%%
%% Bibliography
%%

%% Please use bibtex,

\bibliography{lipics-v2021-article}

\end{document}